\documentclass[iicol,pdflatex,sn-mathphys-num]{sn-jnl}% Math and Physical Sciences Numbered Reference Style
\usepackage{graphicx}%
\usepackage{multirow}%
\usepackage{amsmath,amssymb,amsfonts}%
\usepackage{amsthm}%
\usepackage{mathrsfs}%
\usepackage[title]{appendix}%
\usepackage{xcolor}%
\usepackage{textcomp}%
\usepackage{manyfoot}%
\usepackage{booktabs}%
\usepackage[normalem]{ulem}
\usepackage{algorithm}%
\usepackage{algorithmicx}%
\usepackage{algpseudocode}%
\usepackage{listings}%
\usepackage{lineno}

\usepackage{booktabs}
\usepackage{siunitx}
\usepackage{caption}

\theoremstyle{thmstyleone}%
\theoremstyle{thmstyletwo}%

\theoremstyle{thmstylethree}%

\newcommand{\hl}{$\rm ^4_{\Lambda}H$}

\newcommand{\hel}{$\rm ^4_{\Lambda}He$}

\newcommand{\sNN}[1]{$\sqrt{s_\mathrm{NN}}$~}
\newcommand{\bl}{$B_{\Lambda}$}
\newcommand{\la}{$\Lambda$}
\newcommand{\dblg}{$\Delta B_{\Lambda}^{4}(0^+)$}
\newcommand{\dble}{$\Delta B_{\Lambda}^{4}(1^+)$}

\begin{document}
%\linenumbers

%\title{Observation of Symmetric Energy-Level Splitting from Charge-Symmetry Breaking in $A$~=~4 Mirror Hypernuclei}

\title{Measurement of energy-level splitting from Charge-Symmetry Breaking in $A$ = 4 mirror hypernuclei}

%\author[1,*]{Tianhao Shao}
%\affil[1]{Affiliation, department, city, postcode, country}
\author{The STAR Collaboration}
%\affil[*]{corresponding.author@email.example}

%\affil[+]{these authors contributed equally to this work}

%\keywords{Keyword1, Keyword2, Keyword3}

\abstract{
\unboldmath
Breaking of fundamental symmetries is a ubiquitous phenomenon in physics, underlying the origin of mass and the emerging structure in the universe. The charge symmetry of $\Lambda$ hyperon–nucleon interactions can be probed through the difference in the $\Lambda$ binding energy ($B_{\Lambda}$) between mirror hypernuclei.
In this paper, the \bl~of mirror hypernuclei with atomic mass number $A$ = 4, \hl~and \hel, are measured in Au+Au collisions at the center-of-mass energy of $\sqrt{s_{\rm NN}}$~=~3~GeV with the STAR experiment at RHIC. For the ground states, we obtain 
\bl(\hl) = 2.24 ± 0.02 (stat.) ± 0.04 (syst.) MeV and \bl(\hel) = 2.39 ± 0.05 (stat.) ± 0.05 (syst.) MeV, yielding a charge-symmetry breaking (CSB) effect at the level of 0.15 ± 0.05 (stat.) ± 0.04 (syst.) MeV. In combination with previous measurements of $\gamma$-ray transitions from their $1^+$ excited states, the CSB in excited states is determined to be $-$0.17 ± 0.05 (stat.) ± 0.04 (syst.) MeV. These measurements provide a precise determination of CSB in the hypernuclear system, and establish that the \la~binding energy differences in ground and excited states are comparable in magnitude but opposite in sign, offering new insight to the CSB effect in $\Lambda$–nucleon interactions.
}

%\keywords{Charge symmetry breaking, hypernuclei, YN interaction, heavy-ion collisions}

\maketitle

%\section{Introduction}\label{sec1}

In nuclear physics, charge symmetry~\cite{Coon:1987rhoomega,Miller:1990iz} pertains to the invariance of the nuclear force under interchange of proton and neutron. It is a fundamental feature of the strong interaction and closely related to the isospin symmetry~\cite{Heisenberg:1932,Wigner:1937} in which the proton and neutron are two states of the nucleon ($N$) with isospin $I={1\over 2}$. If charge symmetry were exact, a pair of mirror nuclei, whose proton and neutron numbers are swapped, would exhibit identical binding energies once electromagnetic contributions are removed.
Deviations from this expectation provide a direct probe of charge symmetry breaking (CSB). A well-known example arises from triton ($t$) and 
$^3$He, whose binding energy difference after subtracting the Coulomb contribution is $B(^3{\rm He})-B(t)$~$\approx$~67~$\pm$~9~keV~\cite{Machleidt:2000vh}, a clear manifestation of CSB. This CSB in nucleon-nucleon ($NN$) interactions can be partially attributed to the mixing of the rho and omega measons, originating from the mass difference between the up and down quarks~\cite{Coon:1987rhoomega,Machleidt:2000vh}. 

The \la~hypernuclei provide a unique laboratory where the role of CSB in the \la–nucleon (\la$N$) interaction can be directly investigated~\cite{Hashimoto:2006aw,STAR:2019wjm,Haidenbauer:2025zrr,Coon:1998jd}, as they contain both nucleons and a \la~hyperon.
The \la~binding energy (\bl, also called \la~separation energy) of a hypernucleus is defined as:
\begin{align}
    \label{eq1}
    B_{\Lambda}~=~(m_{\rm core}~+~m_{\Lambda}~-~m_{\rm hyp})c^2,
\end{align}
in which the $m_{\rm core}$, $m_{\Lambda}$, and $m_{\rm hyp}$ are the masses of the nucleon core, the \la~hyperon, and the hypernucleus, respectively. This observable directly relates to the \la$N$ interaction and provides constraints on hyperon-nucleon ($YN$) interactions that enter neutron star equation-of-state calculations~\cite{Fortin:2017cvt}. 
For the pair of mirror hypernuclei with atomic mass number $A$ = 4, \hl~and \hel, whose nucleon cores are $t$ and $^3$He, respectively, they should have nearly the same \bl~under charge symmetry, since the charge-neutral $\Lambda$ hyperon minimizes the Coulomb effect~\cite{Bodmer:1985km}. 

In nuclear emulsion experiments performed in the 1970s, the \bl~difference between \hl~and \hel, defined as \bl(\hel)~$-$~\bl(\hl), was measured to be \dblg~=~350~$\pm$~50~keV for ground states ($0^+$)~\cite{Juric:1973zq}. At that time, the energies of $\gamma$ rays emitted from the transition of their excited states, \hl~($1^+$) and \hel~($1^+$), to ground states were measured to be $1.09\pm0.03~{\rm MeV}$~\cite{Bedjidian:1976zh} and $1.15\pm0.04~{\rm MeV}$~\cite{CERN-Lyon-Warsaw:1979ifx}, respectively, yielding a similar \bl~difference in excited states \dble~=~280~$\pm$~60~keV~as \dblg.
In 2015, the J-PARC E13 Collaboration performed an unambiguous, high-precision measurement of the $\gamma$-ray transition energy of \hel~($1^+$)~to be 1.406 $\pm$ 0.003~MeV~\cite{J-PARCE13:2015uwb}, superseding the earlier measurement~\cite{CERN-Lyon-Warsaw:1979ifx}. Combined with earlier results~\cite{Juric:1973zq,Bedjidian:1976zh}, \dble~was updated to 30~$\pm$~50~keV. This is much smaller than that for the ground states, suggesting a spin dependence of CSB~\cite{J-PARCE13:2015uwb}. In 2016, the A1 Collaboration at the Mainz Microtron measured the \bl~of ground state \hl~with decay-pion spectroscopy~\cite{A1:2015isi,A1:2016nfu}, which prompted the updated values: \dblg~=~233~$\pm$~92~keV and \dble~=~-83~$\pm$~94~keV. This update still favors that the CSB in excited states is smaller than that in ground states. A CSB effect of this magnitude has posed a long‑standing puzzle in our understanding of the $\Lambda N$ interaction for nearly five decades~\cite{Davis:2005mb}. It is worth noting that the systematic uncertainties of early emulsion experiments may be underestimated by 40 to 150~keV~\cite{Gajewski:1967ruj,Davis:2005mb,A1:2015isi}. 

The STAR experiment at the Relativistic Heavy Ion Collider collected data from Au+Au collisions at the nucleon-nucleon center-of-mass energy of $\sqrt{s_{\rm NN}}$~=~3 GeV in the years 2018 and 2021. These collisions are expected to produce large yields of hypernuclei~\cite{STAR:2021orx}. A pilot measurement of CSB in $A$ = 4 hypernuclei, based on approximately 300 million collision events from the 2018 dataset, was published in 2022~\cite{STAR:2022zrf}. The results of \dblg~=~160~$\pm$~140(stat.)~$\pm$~100(syst.)~keV and \dble~=~-160~$\pm$~140(stat.)~$\pm$~100(syst.)~keV suggest \dble$\approx-$\dblg~$<$~0. However, the large statistical and systematic uncertainties prevented a definitive conclusion~\cite{STAR:2022zrf}. Precise experimental measurements of the $A=4$ hypernuclear system are desired to clarify this feature of CSB~\cite{Chen:2025eeb}.

Measurements of $\Lambda$ binding energies for hypernuclei also provide important inputs for astrophysical studies. For example, the authors of Refs.~\cite{Lonardoni:2013gta,Lonardoni:2014bwa} calculated the equation of state and the mass–radius relation of neutron stars by introducing a $\Lambda NN$ three-body force, constrained by the measured $\Lambda$ binding energies of hypernuclei. They demonstrated that different experimental constraints, particularly those from light hypernuclei,significantly affect the predicted neutron star mass~\cite{Lonardoni:2013gta,Lonardoni:2014bwa}.

%\section{Experiment}
In this paper, we present a new high-precision measurement based on a large dataset of approximately 2 billion events collected in 2021. The experimental setup consists of a stationary gold target with 0.25~mm thickness mounted inside the beam pipe and approximately 2~m away from the center of the STAR Time Projection Chamber (TPC)~\cite{Anderson:2003ur}. A gold beam with an energy of 3.85~GeV per nucleon hits the gold target with the produced particles flying towards the STAR detector. \hl~and \hel~produced in collisions are reconstructed through their 2-body, \hl~$\to {\rm ^4He}+\pi^-$, and 3-body, \hel~$\to {\rm ^3He}+p+\pi^-$, decay channels, respectively. Identification of decay daughters is achieved primarily by measuring their mean ionization energy loss per unit length $\langle dE/dx\rangle$ in the TPC. A solenoidal magnetic field of 0.5 T, aligned along the beam direction, enables the determination of particle momenta based on the curvature of their helical tracks in the TPC. Additionally, velocity measurements based on the Time-of-Flight (TOF) detector~\cite{Llope:2012zz,Chen:2024aom} are used to identify particle species when they have left signals in TOF.

The decay topology of the parent hypernucleus is reconstructed from the identified decay daughters by using the KFParticle~\cite{Ju:2023xvg} package based on the Kalman filter. The covariance matrices of momenta and spatial coordinates of tracks are used to calculate the $\chi^2$ between two tracks or one track to the collision vertex at the points of distance of closest approach (DCA). Selections based on the $\chi^2$ of different DCAs are applied to increase the significance of the reconstructed signal. For the \hel~$\to {\rm ^3He}+p+\pi^-$ decay, a cut on $\chi^2$ between the $p+\pi^-$ pair and the collision vertex is introduced to suppress contamination from $\Lambda$ random coincidences. The hypernucleus invariant mass is calculated as $m_{\rm hyp}=\sqrt{(\sum E_{i})^2-(\sum \vec{p}_{i})^2}$, in which $E_{i}$ and $\vec{p}_{i}$ are the energy and momentum of the $i^{\rm th}$ decay daughter. Due to the imperfect reconstruction of the track momentum, a scale factor of 99.8\%, determined by matching the reconstructed mass of $K_{\rm S}^{0}\to\pi^++\pi^-$ in data to the PDG~\cite{ParticleDataGroup:2024cfk} value, is applied to the decay daughters. In addition, while the track energy loss in the detector materials has been corrected during data production assuming pion mass, further energy-loss corrections are needed for heavier $^4$He, $^3$He, and proton, which are evaluated using Monte Carlo simulations of the STAR detector (see Methods for more details). With these corrections applied to the decay daughters, the measured mass of $\Lambda\to p+\pi^-$ in data is consistent with its PDG~\cite{ParticleDataGroup:2024cfk} value within 10~keV. Corrections to particle momenta are performed as a function of particle rapidity in addition to momentum dependence. This reduces systematic uncertainties compared to the previous STAR measurement~\cite{STAR:2022zrf}.

\begin{figure}
    \centering
    \includegraphics[width=1.0\linewidth]{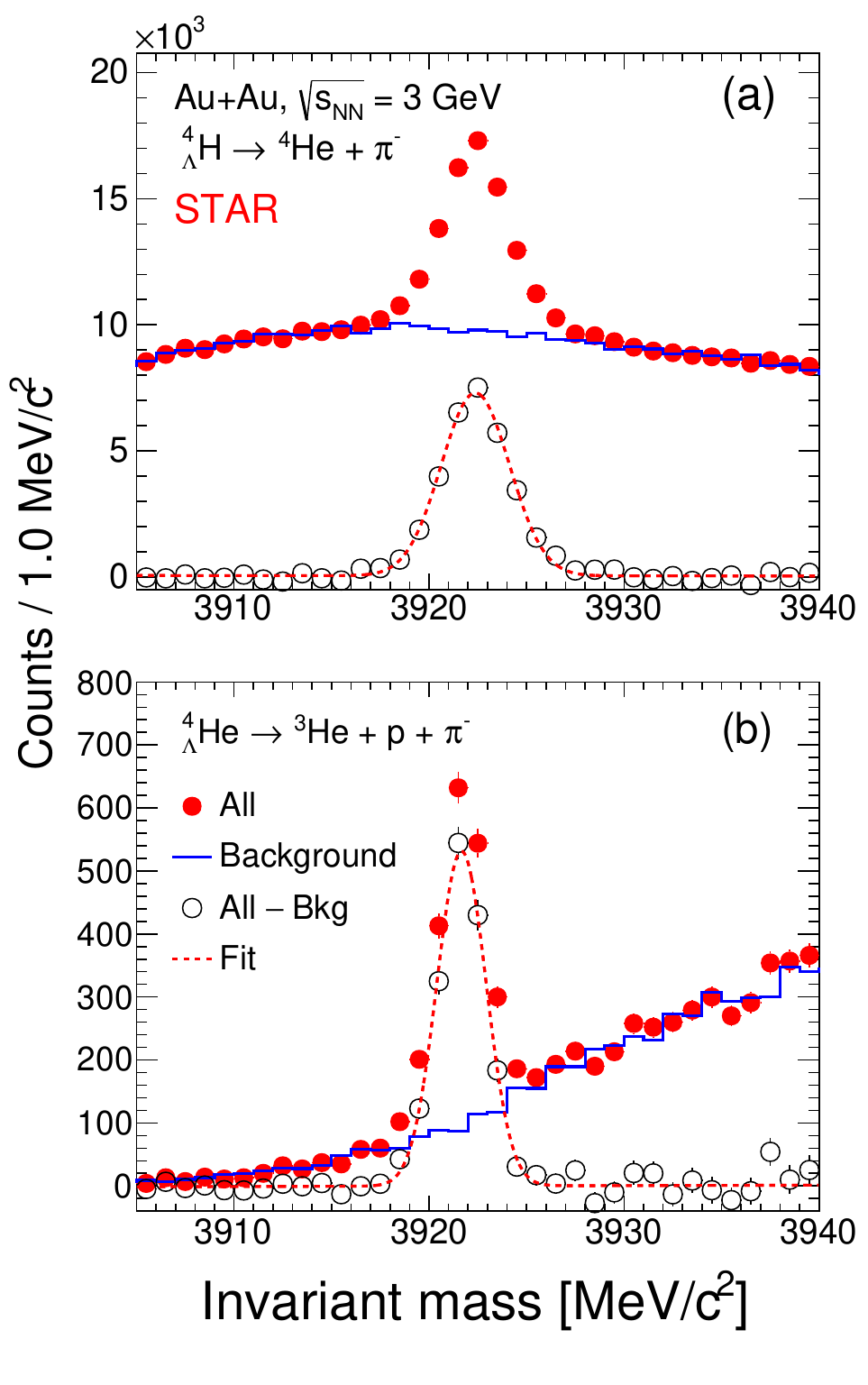}
    \caption{\textbf{The reconstructed invariant mass distributions of \hl~(a) and \hel~(b)}. Solid circles and histograms represent distributions for all candidates (All) and background (Bkg), respectively. Open circles represent the extracted signal distributions (All $-$ Bkg) obtained by subtracting the background distributions from those of all candidates. Dashed curves show the fits to the All - Bkg distributions with a Gaussian function plus a linear function to describe the signal and the residual background. The error bars represent statistical uncertainties.}
    \label{fig:invmass}
\end{figure}

Figure~\ref{fig:invmass} shows the reconstructed invariant mass distributions of \hl~(a) and \hel~(b). Since the excited states first transit to ground states which decay subsequently, the reconstructed signals are for the ground states.
\begin{figure*}
    \centering
    \includegraphics[width=0.8\linewidth]{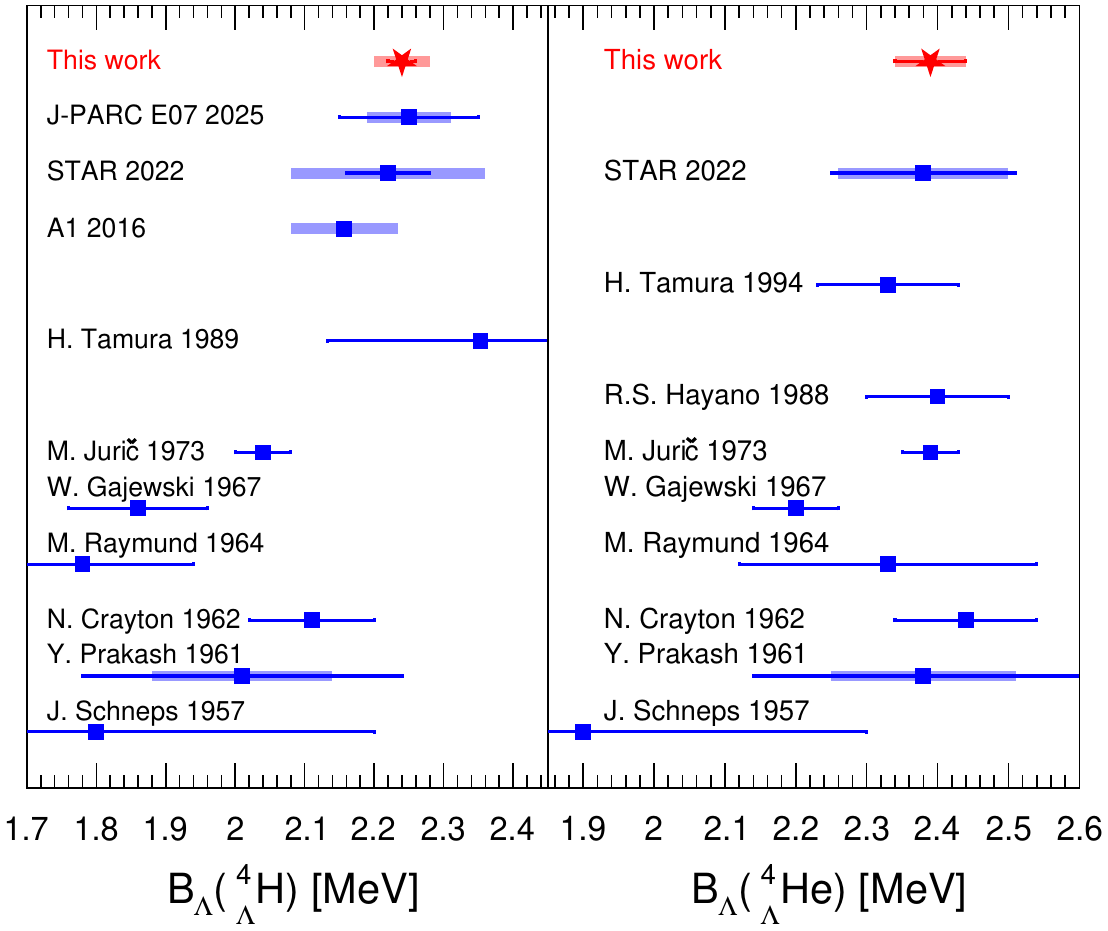}
    \caption{\textbf{\la~binding energies of ground-state \hl~(left) and \hel~(right) from various experiments~\cite{Schneps1957,Prakash1961,Crayton1962,Raymund1964,Gajewski:1967ruj,Juric:1973zq,Tamura:1988kk,Tamura1994,Hayano:1988pn,A1:2016nfu,STAR:2022zrf,Kasagi:2025mvh}.} The error bars and boxes show the statistical and systematic uncertainties. Some experiments only provided statistical uncertainties. The statistical uncertainty of A1 2016 is smaller than the marker size.}
    \label{fig:bindingenergy}
\end{figure*}
\begin{figure}
    \centering
    \includegraphics[width=1.0\linewidth]{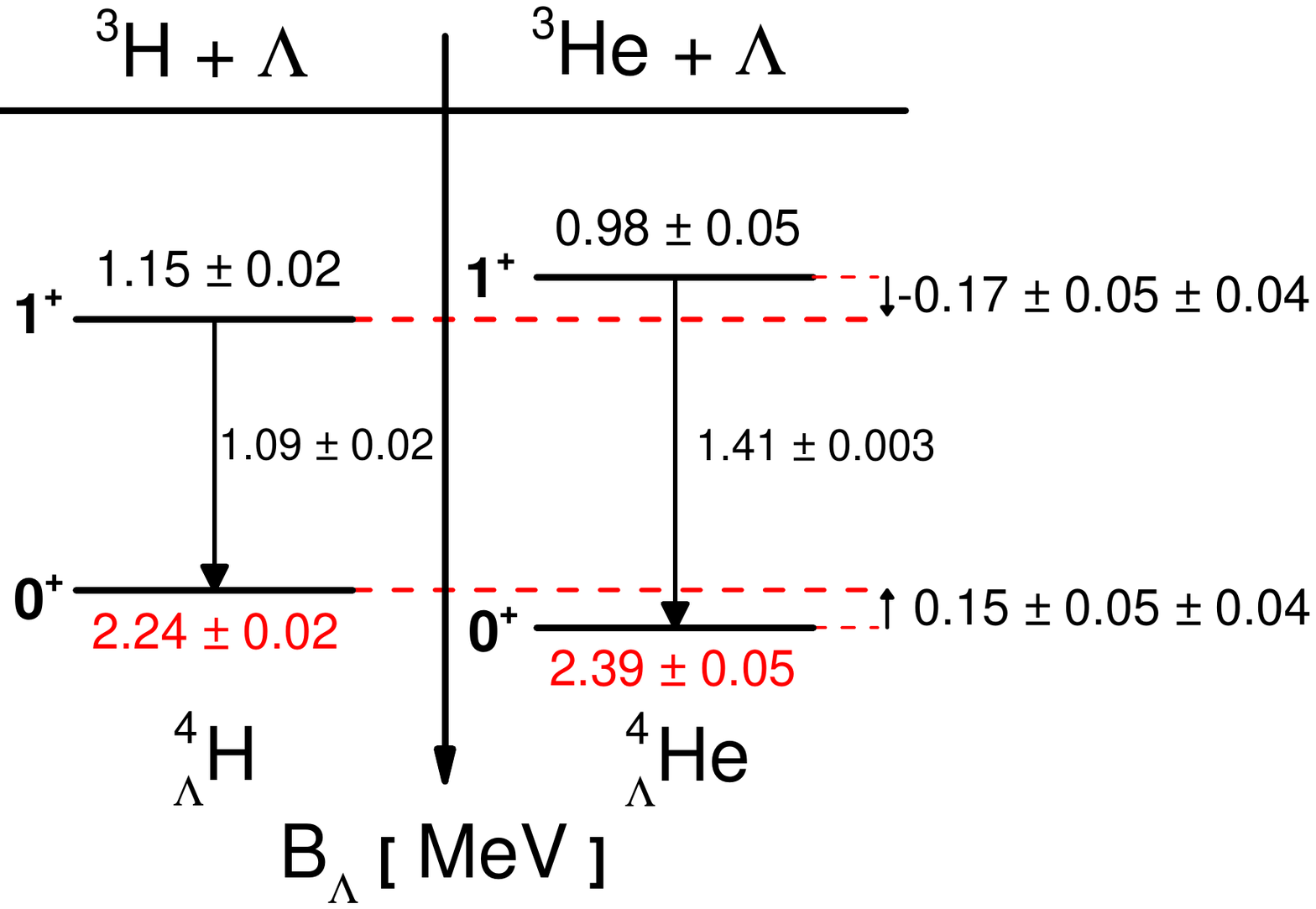}
    \caption{\textbf{Energy level schemes in terms of \la~binding energies of \hl~and \hel.} The values of ground states come from this measurement while the excited states are calculated by combining the $\gamma$-ray transition energy measurements from previous experiments~\cite{Bedjidian:1976zh,J-PARCE13:2015uwb}.}
    \label{fig:energylevel}
\end{figure}
The combinatorial background, shown as histograms and labelled Bkg, is constructed by randomly rotating the tracks of $^4$He or $^3$He particles. This azimuthal rotation disrupts inter-particle correlations while preserving the overall kinematic distributions, thereby providing an accurate estimate of the combinatorial background. The open circles in Fig.~\ref{fig:invmass} denote the invariant mass distributions obtained by subtracting the Bkg distributions from those of all candidates (All). The significances of the signals $S/\sqrt{S+B}$, where $S$ and $B$ are the counts of (All $-$ Bkg) and Bkg distributions in the signal window of (3.916, 3.927)~GeV/$c^2$, are 86 for \hl~and 31 for \hel. By fitting the (All $-$ Bkg) distribution with a Gaussian function describing the shape of the signal plus a linear function for the remaining residual background after the subtraction, the mass values of \hl~and \hel~are taken as the mean of the Gaussian fits. The usage of the Gaussian distribution for describing the signal shape is motivated by the dominance of the detector resolution in the width of the signal peak over the intrinsic decay width. The final \hl~and \hel~masses are obtained by taking the average values of measured \hl~and \hel~masses in different transverse momentum $p_{\rm T}$ intervals, where the inverses of the statistical uncertainties squared are used as weights. Specifically: 
{
\setlength{\abovedisplayskip}{5pt}  
\setlength{\belowdisplayskip}{5pt}
\begin{align*}
&m_{_{\Lambda}^{4}{\rm H}}~=~3922.36\pm0.02({\rm stat.})\pm0.04({\rm syst.})~{\rm MeV}/c^2, \\
&m_{_{\Lambda}^{4}{\rm He}}~=~3921.68\pm0.05({\rm stat.})\pm0.05({\rm syst.})~{\rm MeV}/c^2. 
\end{align*}
}Systematic uncertainties arise from the imperfections in the correction factors for the track momentum and the track energy loss, and the variations in decay topology selection criteria (see Methods for more details).

Combining the masses of $\Lambda$ from PDG~\cite{ParticleDataGroup:2024cfk} ($m_{\Lambda}=1115.683\pm0.006$~MeV/$c^2$), nucleon cores from CODATA~\cite{CPC41} ($m_{t}=2808.92$~MeV/$c^2$, $m_{{\rm ^3He}}=2808.39$~MeV/$c^2$, with negligible uncertainties), the \bl~of ground-state \hl~and \hel~are determined to be
{
\setlength{\abovedisplayskip}{5pt} 
\begin{align*}
&B_{\Lambda}({_{\Lambda}^{4}{\rm H}})~=~2.24\pm0.02~({\rm stat.})\pm0.04~({\rm syst.})~{\rm MeV}, \\
&B_{\Lambda}({_{\Lambda}^{4}{\rm He}})~=~2.39\pm0.05~({\rm stat.})\pm0.05~({\rm syst.})~{\rm MeV}.
\end{align*}
}

Figure~\ref{fig:bindingenergy} summarizes the measured \bl~of ground-state \hl~and \hel~from various experiments. This new measurement of the ground-state $B_{\Lambda}({_{\Lambda}^{4}{\rm H}})$ from the STAR Collaboration deviates from the nuclear emulsion experiment in 1973~\cite{Juric:1973zq} by about 3$\sigma$. It was suggested in Ref.~\cite{Liu:2019mlm} that such a difference could be largely accounted for by the different nucleon core and $\Lambda$ masses used in 1970s and this work. In contrast, the current result is consistent with all measurements after 1973, while providing significantly improved precision. Notably, J-PARC E07 recently measured the \bl~of the ground-state \hl~in nuclear emulsions with deep-learning analyses~\cite{Kasagi:2025mvh}. Their result of $B_{\Lambda}({_{\Lambda}^{4}{\rm H}})=2.25\pm0.10~({\rm stat.})\pm0.06~({\rm syst.})~{\rm MeV}$ is consistent with the STAR measurement with a factor of 5 larger statistical uncertainty.

The $\Lambda$ binding-energy differences between \hl~and \hel~in their ground states can be obtained according to the measured \bl:
{
\setlength{\abovedisplayskip}{5pt}  
\setlength{\belowdisplayskip}{5pt}  
\begin{align*}
    \Delta B_{\Lambda}^{4}(0^+)~&=~0.15\pm0.05~({\rm stat.})\pm0.04~({\rm syst.})~{\rm MeV}.
\end{align*}
}
The systematic uncertainties from track momentum and energy loss corrections cancel in the calculation of the binding-energy difference, while other sources are added in quadrature. 

The $\Lambda$ binding energies of the excited \hl~and \hel~states are determined by combining these results with the $\gamma$-ray transition energies measured from previous experiments: $E_{\gamma}({_{\Lambda}^{4}{\rm H}})=1.09\pm0.02~{\rm MeV}$ (weighted average value of 1.09$\pm$0.03~MeV~\cite{Bedjidian:1976zh}, 1.04$\pm$0.04~MeV~\cite{CERN-Lyon-Warsaw:1979ifx}, and 1.114$\pm$0.030~MeV, as suggested by Ref.~\cite{J-PARCE13:2015uwb}) and $E_{\gamma}({_{\Lambda}^{4}{\rm He}})=1.41\pm0.003~{\rm MeV}$~\cite{J-PARCE13:2015uwb}. Consequently, the $\Lambda$ binding-energy difference in their excited states is obtained:
{
\setlength{\abovedisplayskip}{5pt}  
\setlength{\belowdisplayskip}{5pt}  
\begin{align*}
    \Delta B_{\Lambda}^{4}(1^+)~&=~-0.17\pm0.05~({\rm stat.})\pm0.04~({\rm syst.})~{\rm MeV}.
\end{align*}
}The measured \bl~of \hl~and \hel~in their ground and excited states, together with the differences between them, which differ from 0 by 2.3$\sigma$ and 2.6$\sigma$, respectively, are illustrated in Fig.~\ref{fig:energylevel}.

\begin{figure*}[!htbp]
    \centering
    \includegraphics[width=0.8\textwidth]{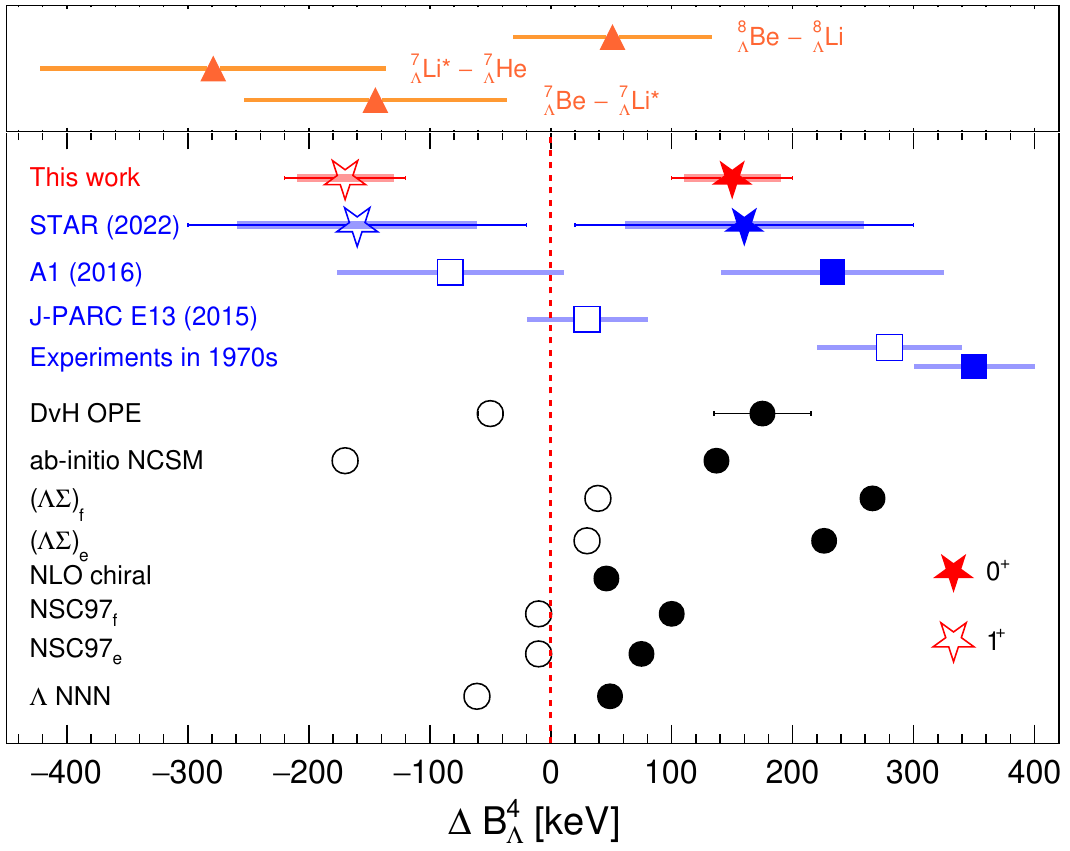}
    \caption{\textbf{The \la~binding energy differences between \hl~and \hel~in their ground states (solid markers) and excited states (open markers).} The red and blue stars represent the results from this measurement and the previous work using the STAR 2018 data~\cite{STAR:2022zrf}. Error bars show their statistical uncertainties and boxes show the systematic uncertainties. The blue squares represent the previous results~\cite{Juric:1973zq,A1:2015isi,A1:2016nfu,CERN-Lyon-Warsaw:1979ifx,Bedjidian:1976zh,J-PARCE13:2015uwb} and the shaded boxes show their total uncertainties. J-PARC E13 only measured \hel($1^+$) and the gorund-state \bl~were taken from the nuclear emulsion experiment~\cite{Juric:1973zq}. The black circles represent calculations with different theoretical models~\cite{Gazda:2015qyt,Gazda:2016qva,Coon:1998jd,Nogga:2001ef,Haidenbauer:2007ra,Nogga:2013pwa,Gal:2015bfa} and the error bars show their uncertainties. The red dashed vertical line represents $\Delta B_{\Lambda}^4=0$. Upper panel: orange triangles show the measured \la~binding-energy differences of the $A$ = 7 and 8 hypernuclear isospin multiplets ($_{\Lambda}^7{\rm Li}^*$ is the excited isospin $I$~=~1 state of $_{\Lambda}^7{\rm Li}$)~\cite{HypernuclearDataBase,Haidenbauer:2025zrr}, with shaded boxes indicating total uncertainties.}
    \label{fig:CSB_compare}
\end{figure*}

%\section{Discussion}
Figure~\ref{fig:CSB_compare} summarizes the $\Lambda$ binding energy differences from this work, together with the previous STAR measurement~\cite{STAR:2022zrf}, other previous measurements~\cite{Juric:1973zq,A1:2015isi,A1:2016nfu,CERN-Lyon-Warsaw:1979ifx,Bedjidian:1976zh,J-PARCE13:2015uwb}, and theoretical calculations~\cite{Gazda:2015qyt,Gazda:2016qva,Coon:1998jd,Nogga:2001ef,Haidenbauer:2007ra,Gal:2015bfa}.
Owing to the substantially larger data sample and the implementation of more refined correction procedures, the present results have significantly reduced statistical and systematic uncertainties, about 63\% in total uncertainty, relative to previous STAR measurements~\cite{STAR:2022zrf}. The combined results of the previous and current measurements are described in Methods. The new results also have much better precision than measurements of the $\Lambda$ binding-energy differences between the $A$~=~7 isospin triplets [$_{\Lambda}^7$He, $_{\Lambda}^7{\rm Li}^*$ (the excited isospin $I$~=~1 state of $_{\Lambda}^7{\rm Li}$), and $_{\Lambda}^7$Be)], $A$~=~8 hypernuclei ($_{\Lambda}^8$Li and $_{\Lambda}^8$Be)~\cite{HypernuclearDataBase,Haidenbauer:2025zrr} shown in the upper panel of Fig.~\ref{fig:CSB_compare}, and heavier hypernuclei whose uncertainties are all larger than 200~keV~\cite{HypernuclearDataBase}.

The new STAR measurements indicate that the $\Lambda$ binding-energy differences between \hl~and \hel~in their ground and excited states have similar magnitudes with opposite signs (\dble$\approx-$\dblg~$<$~0), suggesting a symmetric energy-level splitting from the CSB effect (see Methods for more details). In quantum mechanics, when a symmetry is broken in a system that has only two separated energy levels with a double-well potential, or equivalently, when an off-diagonal perturbation is added to a Hamiltonian diagonalized in a two-dimensional space, the resulting energy shifts of the two energy levels should have opposite signs and similar magnitudes~\cite{gilmore2004elementary}. Applying this concept to the $A=4$ hypernuclear system, the \la~binding energy shifts of \hel~relative to \hl~in their ground and excited states should exhibit opposite signs and similar magnitudes under CSB. The new STAR measurements align with this expectation. 

Several theoretical predictions of \la~binding-energy differences for both ground and excited states are based on \la$\Sigma$ mixing and the mass difference between \la~and $\Sigma$, as suggested by Dalitz and Von Hippel~\cite{Dalitz:1964es}. These predictions are shown as black circles in Fig.~\ref{fig:CSB_compare} for comparison. They are sorted according to their published dates starting from the bottom. The $\Lambda NNN$ four-body calculation~\cite{Coon:1998jd} supports the symmetric energy-level splitting. However, it seems to underestimate the magnitudes even with charge asymmetric $\Lambda N$ interactions. Theoretical works from Refs.~\cite{Nogga:2001ef,Haidenbauer:2007ra} used $\Lambda\Sigma$ coupling potentials from NSC97$_{\rm e/f}$ models in their four-body calculations. They show much larger \dblg~than \dble~and they have opposite signs. A four-body calculation with NLO chiral $\Lambda N$ interaction predicts a small \dblg~\cite{Nogga:2013pwa}. Calculations introducing one-pion exchange by allowing $\Lambda \Sigma$ mixing in SU(3) (labelled $(\Lambda\Sigma)_{\rm e/f}$) suggest much larger \dblg~than \dble~with the same sign~\cite{Gal:2015bfa}. None of the NSC97$_{\rm e/f}$ or $(\Lambda\Sigma)_{\rm e/f}$ models can describe the STAR measurements since the $\Sigma N$ admixture in $1^+$ states is considered to be weaker than in $0^+$ states~\cite{Gal:2015bfa}. The calculation of $ab$-$initio$ no-core shell model with $YN$ potentials from Bonn–Jülich leading-order chiral effective field theory plus a CSB $\Lambda$-$\Sigma^0$ mixing vertex~\cite{Gazda:2015qyt} (labelled $ab$-$initio$ NCSM and the results are shown with parameters of $\Lambda_{\rm cutoff}$~=~600~MeV and $\hbar \omega$~=~26~MeV) describes STAR measurements quantitatively. However, in the updated calculation~\cite{Gazda:2016qva} with the Dalitz–von Hippel one-pion exchange (DvH OPE) mechanism~\cite{Dalitz:1964es}, the \dble~becomes smaller and underestimates the STAR results. This is due to that the one-pion exchange in the $ab$-$initio$ NCSM contributes oppositely in sign relative to the DvH OPE~\cite{Gazda:2016qva}. 

The new STAR measurements presented here achieve substantially higher precision than those obtained from the 2018 data~\cite{STAR:2022zrf}, providing strong inputs for theoretical studies. As suggested in Ref.~\cite{Haidenbauer:2021wld}, the measured CSB in \hl~and \hel~can be used to constrain the scattering length of the $\Lambda$-neutron interaction, which is difficult to measure directly by scattering experiment. A recent review~\cite{Haidenbauer:2025zrr} emphasizes that the approach used in Ref.~\cite{Haidenbauer:2021wld} should be repeated with new data, with proper consideration of experimental uncertainties. The measured CSB can also be used to predict the $\Lambda$ binding energies and CSB of heavy hypernuclei. The authors of Ref.~\cite{Le:2022ikc} and Ref.~\cite{Haidenbauer:2025zrr} calculated the CSB in $A$~=~7 and 8 hypernuclei with an $ab$-$initio$ no-core shell model plus CSB $\Lambda N$ potential constrained by measured $\Lambda$ binding energies and their difference in \hl~and \hel. With the constraints based on the central values of \dblg~and~\dble~from the previous STAR measurement using 2018 data~\cite{STAR:2022zrf}, the predicted CSB in $A$~=~7 and 8 hypernuclei describe the experimental results better than being constrained by the A1 experiment~\cite{A1:2016nfu}. In a recent theoretical work~\cite{Le:2024rkd}, the authors extracted and constrained the $YNN$ interaction according to the \bl~of $A$~=~4 hypernuclei and described the energy levels of $^7_{\Lambda}$Li successfully.

The improved $^4_\Lambda\mathrm{H}$ and $^4_\Lambda\mathrm{He}$ binding energies also place tighter bounds on the spin-dependent components of CSB that underpin modern descriptions of light hypernuclei~\cite{Haidenbauer:2025zrr}. In doing so, they reduce longstanding model ambiguities associated with $\Lambda$-$\Sigma$ mixing~\cite{Gal:2015xra,Haidenbauer:2013dua}, the interplay of two- and three-body hyperon–nucleon forces~\cite{Lonardoni:2013gta,Lonardoni:2014bwa,Contessi:2019csf}, and the contact versus spin-dependent contributions required to reproduce the $A=4$ level structure~\cite{Gazda:2015qyt,Gazda:2016qva}. As the $YNN$ interaction inferred from light-hypernuclear separation energies also provides the repulsive contribution needed to counteract the softening induced by the inclusion of the $\Lambda$ hyperons at the high densities, such as inside neutron stars, the present results further restrict the range of $Y N$ and $Y NN$ strengths consistent with few-body data. Together, these advances strengthen the empirical foundation for assessing hyperon effects in dense matter and sharpen the constraints relevant to proposed resolutions of the hyperon puzzle~\cite{Lonardoni:2013gta,Lonardoni:2014bwa,Chatterjee:2015pua,Ghosh:2022NS}. 

\section*{Acknowledgment}
We thank the RHIC Operations Group and SDCC at BNL, the NERSC Center at LBNL, and the Open Science Grid consortium for providing resources and support.  This work was supported in part by the Office of Nuclear Physics within the U.S. DOE Office of Science, the U.S. National Science Foundation, National Natural Science Foundation of China, Chinese Academy of Science, the Ministry of Science and Technology of China and the Chinese Ministry of Education, NSTC Taipei, the National Research Foundation of Korea, Czech Science Foundation and Ministry of Education, Youth and Sports of the Czech Republic, Hungarian National Research, Development and Innovation Office, New National Excellency Programme of the Hungarian Ministry of Human Capacities, Department of Atomic Energy and Department of Science and Technology of the Government of India, the National Science Centre and WUT ID-UB of Poland, German Bundesministerium f\"ur Bildung, Wissenschaft, Forschung and Technologie (BMBF), Helmholtz Association, Ministry of Education, Culture, Sports, Science, and Technology (MEXT), Japan Society for the Promotion of Science (JSPS), and Agencia Nacional de Investigacion y Desarrollo de Chile (ANID), Chile.





\bibliography{mybib}% common bib file

\clearpage
\appendix
\section{Methods}
\begin{figure*}
    \centering
    \includegraphics[width=1.0\linewidth]{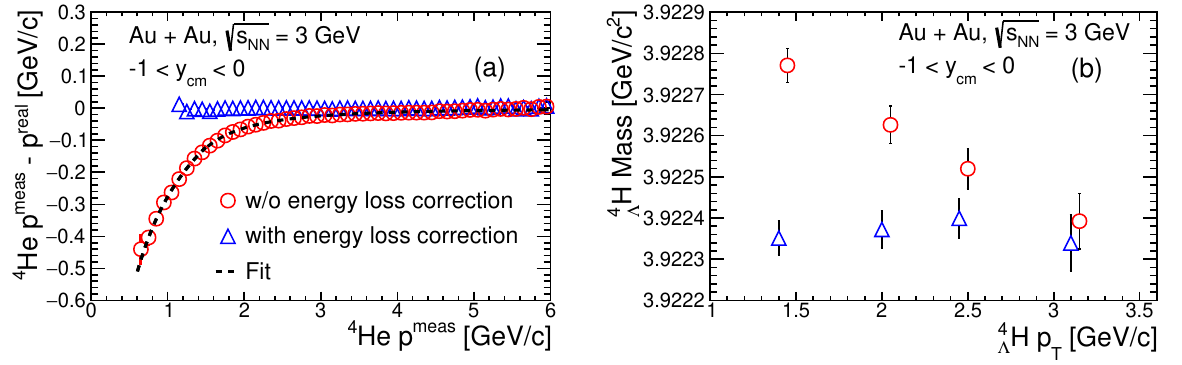}
    \caption{\textbf{Left panel (a): The average momentum shift $p^{\rm meas}$ - $p^{\rm real}$ as a function of the measured momentum $p^{\rm meas}$ for $^4$He in the center-of-mass rapidity region $-1 < y_{\rm cm} < 0$.} The circles and triangles represent results without and with energy loss corrections. The dashed curve is the fit to the circles. \textbf{Right panel (b): The measured \hl~mass in different $p_{\rm T}$ regions.} The circles and triangles represent results without and with energy loss corrections. Error bars represent statistical uncertainties.}
    \label{fig:eloss}
\end{figure*}
\subsection{Energy loss correction}
When particles from the collisions travel through the detector materials, they lose energy, resulting in the measured momenta being smaller than the true values. Energy loss corrections for the decay daughters are therefore needed for accurate reconstruction of hypernuclei. During the official data production at STAR, the energy losses of tracks have been corrected by assuming all the tracks are pions. Thus additional energy loss corrections are needed for the momenta of $^4$He, $^3$He, and proton in this work. The correction method is similar to the one used in Ref.~\cite{STAR:2022zrf}. \hl~and \hel~samples, generated by Monte Carlo and propagated through the GEANT3 simulation of the STAR detector geometry and material, are embedded into real data and reconstructed using the same software as data production. The momentum shift of each particle track can be determined by calculating the difference between the measured momentum value $p^{\rm meas}$ and the corresponding truth value $p^{\rm real}$ from Monte Carlo. 

As an example, the open circles in Fig.~\ref{fig:eloss} (a) show the average momentum shift $p^{\rm meas}$ - $p^{\rm real}$ versus $p^{\rm meas}$ of $^4$He in the center-of-mass rapidity region $-1 < y_{\rm cm} < 0$ from simulation. The energy loss effect is significant when the momentum is low, causing the measured \hl~mass in the low transverse momentum ($p_{\rm T}$) region to be larger than that in the high $p_{\rm T}$ region, shown as the circles in Fig.~\ref{fig:eloss} (b).
The open circles in Fig.~\ref{fig:eloss} (a) are fitted with the correction function:
\begin{align}
    p^{\rm meas} - p^{\rm real}~=~\delta_0~+~\delta \left(1~+~\frac{m^2}{(p^{\rm meas})^2} \right)^{\alpha} ,
\end{align}
where $m$ is the mass of $^4$He, and $\delta_0$, $\delta$, and $\alpha$ are fitting parameters. In this work, the momentum shifts are extracted and fit in ten rapidity bins with an equal interval of 0.1 in the region $-1 < y_{\rm cm} < 0$. The $p^{\rm meas}$ of particles in these rapidity regions are corrected with the corresponding correction functions obtained by fitting and then used in the reconstruction of hypernuclei. Shown as the open triangles in Fig.~\ref{fig:eloss} (a), the average momentum shift of $^4$He is approximately zero after energy loss corrections. Consequently, the extracted \hl~mass is independent of \hl~$p_{\rm T}$, as illustrated in Fig.~\ref{fig:eloss} (b).

\subsection{Systematic uncertainties}
The systematic uncertainties of this measurement arise from the corrections on track momentum and energy loss, as well as variations in decay topology selection criteria. 

A conservative absolute uncertainty of $\pm$0.03\% is determined for the correction factor of 99.8\% on the track momentum. This is based on the residual difference of the measured $K_{\rm S}^0$ mass to the PDG value after applying the correction factor as well as its statistical uncertainty. Half of the maximum changes in the measured hypernuclei masses resulting from this source are taken the systematic uncertainties.

In simulations, the extracted masses for $\Lambda$, \hl~and \hel~show almost identical deviations, at the level of 0.03 MeV/$c^2$, to the input values after applying energy loss corrections on proton, $^4$He and $^3$He. This suggests that $\Lambda$ can be used as a benchmark to evaluate the performance of energy loss corrections on the measurements of \hl~and \hel. The measured $\Lambda$ mass in data, after track momentum and energy loss corrections, deviates from the PDG value within $\pm$0.01~MeV/$c^2$. To be conservative, it is propagated to the systematic uncertainties of \hl~and \hel~by scaling it with the ratio of the difference between the measured hypernuclei masses with and without energy loss corrections to this difference of the $\Lambda$ masses.

The systematic uncertainties from decay topology reconstruction are evaluated by varying the selection criteria. Half of the maximum resulting changes in the measured masses are taken as the uncertainties. These systematic uncertainties are summarized in Table~\ref{tab1} and added in quadrature to obtain the total uncertainties. 

The systematic uncertainties from the fit procedure, bin width, and the centrality calibration are found to be negligible.

\begin{table}[!htbp]
    \centering
    \caption{Systematic uncertainties for the masses and \bl~of \hl~and \hel~in MeV.}
    \label{tab1}       % Give a unique label
    % For LaTeX tables you can use
    \setlength{\tabcolsep}{4mm}{
    \begin{tabular}{ccc}
    \hline
    \hline
    Uncertainty source & \hl  & \hel   \\\hline
    Track momentum correction & 0.03 & 0.03 \\
    Energy loss correction & 0.02 & 0.02 \\
    Decay topology reconstruction    &0.02     & 0.03 \\
    Total                  & 0.04       & 0.05       \\
    \hline
    \hline
    \end{tabular}}
    % Or use
    %\vspace*{5cm}  % with the correct table height
\end{table}

For the binding-energy difference, the systematic uncertainties from track momentum and energy loss corrections cancel, while uncertainties from decay topology reconstruction and measurements of $\gamma$-ray energies for excited states~\cite{J-PARCE13:2015uwb} are added in quadrature. Systematic uncertainties for the binding-energy difference are summarized in Table~\ref{tab2}.

\begin{table}[!htbp]
    \caption{Systematic uncertainties for the $\Lambda$ binding-energy difference in ground and excited states in MeV.}
    \label{tab2}
    \centering
    \begin{tabular}{ccc}
        \hline
        \hline
        Uncertainty source   & $\Delta B_{\Lambda}^{4}(0^+)$ & $\Delta B_{\Lambda}^{4}(1^+)$  \\
        %\cline{2-9}% partial hline from column i to column j
        \hline
        \bl(\hl)    & 0.02 & 0.02\\
        \bl(\hel)    & 0.03 & 0.03\\
        $\gamma$ ray    & 0.00 & 0.02\\
        Total & 0.04 & 0.04\\
        \hline
        \hline
    \end{tabular}
\end{table}

\subsection{Combinations of STAR measurements}
The systematic uncertainties of the previous~\cite{STAR:2022zrf} and current STAR measurements that arise from the correction on the track momentum are fully correlated due to similar reconstruction procedure for tracks. Other systematic uncertainties are uncorrelated since the conditions of the STAR detector and data production details are different in the years 2018 and 2021. The combined results of the two STAR measurements are:
\setlength{\abovedisplayskip}{5pt} 
\begin{align*}
B_{\Lambda}({_{\Lambda}^{4}{\rm H}})~&=~2.24\pm0.02({\rm stat.})\pm0.05({\rm syst.})~{\rm MeV}, \\
B_{\Lambda}({_{\Lambda}^{4}{\rm He}})~&=~2.39\pm0.05({\rm stat.})\pm0.05({\rm syst.})~{\rm MeV}, \\
\Delta B_{\Lambda}^{4}(0^+)~&=~0.15\pm0.05({\rm stat.})\pm0.04({\rm syst.})~{\rm MeV}, \\
\Delta B_{\Lambda}^{4}(1^+)~&=~-0.17\pm0.05({\rm stat.})\pm0.04({\rm syst.})~{\rm MeV}.
\end{align*}

\subsection{Symmetry test}
To evaluate the symmetry between \dble~and \dblg~with respect to 0, the value of $R$, which is defined as the ratio of \dble~to \dblg~plus 1, is calculated:
\begin{align}
    R~\equiv~1~+~\frac{\Delta B_{\Lambda}^{4}(1^+)}{\Delta B_{\Lambda}^{4}(0^+)}.
\end{align}
The \dble~and \dblg~are more symmetric when $R$ is closer to 0.

\begin{figure}
    \centering
    \includegraphics[width=1.0\linewidth]{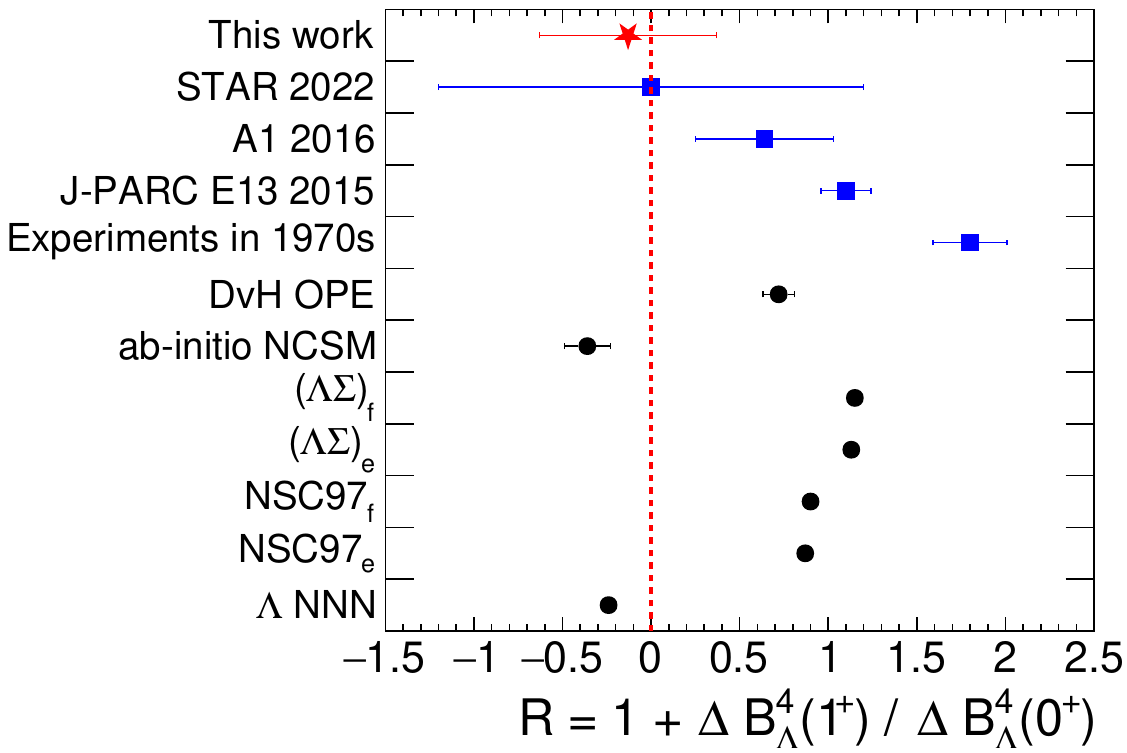}
    \caption{\textbf{The symmetry between \dble~and \dblg~relative to 0 quantified by $R$.} The red star represents the value from this work and the blue squares are from previous experiments~\cite{Juric:1973zq,A1:2015isi,A1:2016nfu,CERN-Lyon-Warsaw:1979ifx,Bedjidian:1976zh,J-PARCE13:2015uwb,STAR:2022zrf}. The black dots are from theoretical calculations~\cite{Gazda:2015qyt,Gazda:2016qva,Coon:1998jd,Nogga:2001ef,Haidenbauer:2007ra,Nogga:2013pwa,Gal:2015bfa}. The error bars represent the total uncertainties. The \dblg~value used for the result labelled ``J-PARC E13 2015'' is the same as for the experiments in the 1970s.}
    \label{fig:symmetry}
\end{figure}

Figure~\ref{fig:symmetry} summarizes the values of $R$ from this work, previous experiments, and various theoretical calculations. Most theoretical calculations deviate from 0, except for the $\Lambda NNN$~\cite{Coon:1998jd} and $ab-initio$ NCSM~\cite{Gazda:2015qyt}. However, as shown in Fig.~\ref{fig:CSB_compare}, the magnitudes of \dble~and \dblg~from $\Lambda NNN$ are smaller than our measurement, while the ones from $ab-initio$ NCSM are similar in magnitude to our measurements as well as having a small and negative R value.

\end{document}